\documentclass[
twocolumn, hf,
]{ceurart}

\usepackage{graphicx}
\usepackage{hyperref}       
\graphicspath{{./images}}
\usepackage{float}
\usepackage{xcolor}
\usepackage{ragged2e}
\usepackage{fixltx2e}
\usepackage{siunitx}
\usepackage{tabularx}
\usepackage{listings}
\usepackage{ragged2e}
\usepackage{arydshln}
\usepackage{indentfirst}
\begin{document}

\copyrightyear{2021}
\copyrightclause{Copyright for this paper by its authors.
  Use permitted under Creative Commons License Attribution 4.0
  International (CC BY 4.0).}

\conference{}

\title{Towards a modeling and analysis environment for industrial IoT systems}

\author[1,2]{Felicien Ihirwe}[%
orcid=0000-0002-4463-6268,
email=felicien.ihirwe@intecs.it,
url=ihirwe.com/,
]
\address[1]{Innovation Technology Services Lab, Intecs Solutions Spa, Pisa, Italy}

\address[2]{Department of Information Engineering Computer Science and Mathematics, University of L'Aquila, L'Aquila, Italy}

\author[2]{Davide Di Ruscio}[%
orcid=0000-0002-5077-6793,
email=davide.diruscio@univaq.it,
url=http://people.disim.univaq.it/diruscio/,
]

\author[1]{Silvia Mazzini}[%
email=silvia.mazzini@intecs.it,
]

\author[2]{Alfonso Pierantonio}[%
orcid=0000-0002-5231-3952,
email=alfonso.pierantonio@univaq.it,
url=https://pieranton.io,
]


\begin{abstract}
The development of Industrial Internet of Things systems (IIoT) requires tools robust enough to cope with the complexity and heterogeneity of such systems, which are supposed to work in safety-critical conditions. The availability of methodologies to support early analysis, verification, and validation is still an open issue in the research community. The early real-time schedulability analysis can help quantify to what extent the desired system's timing performance can eventually be achieved. In this paper, we present CHESSIoT, a model-driven environment to support the design and analysis of industrial IoT systems. CHESSIoT follows a multi-view, component-based modelling approach with a comprehensive way to perform event-based modelling on system components for code generation purposes employing an intermediate ThingML model. To showcase the capability of the extension, we have designed and analysed an Industrial real-time safety use case.
\end{abstract}


\maketitle

\section{Introduction}\label{sec:introduction}
The complexity of industrial IoT systems is deemed to increase due to the growing need for data acquisition, processing, and storage techniques. Moreover, the rapid penetration of advanced machine-to-machine communications in cyber-physical systems triggers even more complexity in IoT production systems \cite{modelingIIoT}. The design complexity of such systems has to consider different layers, including the behaviour of the building components, their inter-connectivity, not to mention the message heterogeneity \cite{asyncAPI,IoTdesign}. 

Model-Driven Engineering (MDE) aims at supporting software development and analysis by promoting the adoption of models as first-class citizens.
Performing early analysis on intended systems can help discover how they will behave once deployed. Furthermore, the quantitative results from the conceptual analysis of such systems can provide theoretical support for optimizing system architectures and parameters earlier enough \cite{iotperformance}. The challenges present in IoT systems validation and verification (V\&V) also poses a significant gap in certifying such systems. Different industrial model-driven approaches such as \cite{uml4iot,MDE4IoT,sysml4IoT,sysml4IoT2} try to cope with such challenges, but we see it as not yet enough.

To cope with the complexity and heterogeneity issues at the levels of system design, operational, and deployment, we propose CHESSIoT. CHESSIoT is a multi-view modelling environment to support the design, development, and analysis of industrial IoT systems. Currently, CHESSIoT is being developed on top of the CHESS tool\cite{CHESScomplex}, a mature modelling and analysis environment for the development of complex industrial systems \cite{CHESSIndustrial}. Since the CHESS tool has been developed to meet industrial needs \cite{lessonCHESS}, we decided to rely on it to support industrial IoT development. In addition to CHESS, CHESSIoT offers an IoT-specific modelling infrastructure where the user can specify the system's structural and behavioural architectures, perform different real-time analyses and generate platform-specific code.

To guaranty a fully decoupled extension, CHESSIoT introduced the \textit{"IoT sub-view"}, and once applied in all design stages, the user will benefit from a dedicated IoT-specific modelling infrastructure consisting of specific diagrams and palettes. The CHESSIoT methodology follows a component-based approach where all system components and their internal behaviours are decomposed separately. The specified components can be annotated with extra-functional properties for analysis purposes or used to generate platform-specific code. The inner or external events can be defined using component state machines. The development of CHESSIoT is still in its early phases but with very significant progresses concerning its design capabilities. 

CHESSIoT will benefit from the system model-based verification and validation resources provided by CHESS. Moreover, CHESSIoT employs ThingML tool \cite{ThingMLcore} to generate code. ThingML is a well-proven software modelling tool aligned with UML (state-charts and components) and an imperative platform-independent action language to construct the intended IoT applications \cite{LCE4IoT}. ThingML model can compile and generate code in different languages such as C/C++, Java, JavaScript, Arduino, and Go.
In this paper, we overview an industrial IoT real-time safety use case to showcase the current capabilities of CHESSIoT. Interested readers can access the complete source code of the developed extension publicly available on GitHub repository.\footnote{\url{https://github.com/feliIhirwe/ChessIoT_Dev}}

The main contributions of this paper can be summarized as follows: 
\begin{itemize}
    \item The CHESSIoT modelling environment to support the development and analysis of industrial IoT systems;
    \item Overview of the envisioned CHESSIoT2ThingML transformation that will enable the generation of target platform code.
    \item Description of a simple use-case on industrial real-time safety to showcase the analysis capabilities for the IoT. 
    \item Showcase the real-time schedulability analysis that can be performed on the CHESSIoT model of the presented use case.
\end{itemize}

This paper is structured as follows: Section \ref{sec:background} gives a concise background of the CHESS and ThingML platforms. Section \ref{sec:CHESSIoT} presents the technical specification of the proposed CHESSIoT extension. Section \ref{sec:usecase} describes the simple industrial use-case. Section \ref{sec:related} examines the related work. Lastly, Section \ref{sec:conclusion} concludes the paper and makes an overview of our future work.

\section{BACKGROUND}\label{sec:background}

The following section gives a brief background on the CHESS tool and the motivation behind its extension. Furthermore, we will briefly present the ThingML tool and its development infrastructure and why it is crucial for the CHESSIoT code generation.

\subsection{CHESS development environment}

The CHESS tool is a mature cross-domain model-driven tool developed on top of the Eclipse Papyrus environment to support the modelling and analysis of dependable systems \cite{CHESScomplex}. The CHESSML modelling language provided by CHESS is an integrated modelling language profiled from OMG standard languages: UML, SysML, and MARTE under the Papyrus modelling environment \cite{papyrus}. The CHESSML language was designed to support the component-based development methodology. Emphasis is given to specify the non-functional properties of the modelled components, including critical properties such as time predictability, isolation, transparency, and other real-time and dependability-related characteristics \cite{chessIoT}.  

The CHESS tool provides a multi-view modelling environment where each view has its own underlined constraints that enforce its specific privileges on model entities and properties that can be manipulated. Depending on the current stage of the design process, CHESS sub-views are adopted to enhance specific design properties or steps of the current process. Different tools, plugins, and languages have been integrated into CHESS to support model validation, model checking, real-time, and dependability analysis.  

CHESS-based modelling and development rely on different views. The \textit{Requirement view} is used to define system requirements and track their verification. The \textit{System view} provides a suitable frame for system-level design activities such as contract-based design. It furthermore reinforces a couple of functional and dependability analyses. The \textit{Component view} consists of two sub-views, i.e., \textit{Functional} and \textit{Extra-functional views} which serve for software design logic between system components and their internal compositions. The goal of the \textit{Deployment view} is to model the hardware structure of the system by permitting the allocation of their corresponding software component instances. The \textit{Analysis view} captures all the activities and diagrams related to the analysis capabilities of CHESS. Finally, the \textit{Instance view} visualizes the resulting instance model generated after performing model-to-model transformations of the software models to hardware allocated components. 

Even though CHESS has been successfully applied in different application domains such as Avionics \cite{chessavionic}, Automotive \cite{chessautomotive}, Space \cite{chessVerification}, Telecommunication \cite{chesstelecom}, and Petroleum \cite{chesspetroleum, FTAandFMEA}, its current status does not explicitly provide modelling capabilities for the IoT domain. Consequently, we aim at extending the existing modelling and analysis infrastructure starting from software modelling infrastructure. 



\subsection{The ThingML framework}

ThingML is amongst the most popular domain-specific model-driven engineering tools for the Internet of Things domain. It comprises a custom textual modelling language, a supporting modelling tool, and advanced code generator capabilities. The ThingML language combines well-proven software modelling constructs aligned with UML (state-charts and components) and an imperative platform-independent action language to construct the intended IoT applications \cite{LCE4IoT}. ThingML code generator targets many popular programming languages such as C/C++, Java, and Javascript, and about ten different target platforms (ranging from tiny 8bit microcontrollers to servers) and ten different communication protocols \cite{ThingMLcore}. 

In ThingML, a Thing can be defined by properties, functions, messages, ports, and a set of state machines. Like in UML, \textit{properties} are local variables to a Thing and can be accessible only to other local behavioural parts of a Thing, such as state machines. Functions are also functional behavioural of a Thing but can also be accessed outside the owner Thing. The interaction with a Thing is enabled through required or provided ports by exchanging messages. A message can have zero or multiple parameters to specify its format. ThingML introduced another kind of a Thing called \textit{fragment} which contains declarations of different Thing's messages. 

ThingML has been used in many different industrial and commercial projects \cite{ThingMLcore}. Several research approaches have been shown interest in applying ThingML as their modelling or code generation framework. To mention a few, in \cite{caps} ThingML has been used to generate code for CAPS, an architecture-driven modelling framework for the development of IoT Systems. In \cite{ENACT} ThingML has been used to specify the behaviour of distributed software components, and later it has been extended with mechanisms to monitor and debug the execution flow of a ThingML program. In \cite{cypriot}, CyprIoT tool used and extended the ThingML modelling language to model the behaviour of IoT things and as a code generator for platform-specific code. 

Although ThingML framework is very mature and looks very promising, it can not be one shoe fit for all aspects that involve IoT systems design and development. For instance, the ThingML framework does not provide the means to conduct any system-related analyses that are very important in the Industrial IoT domain. There is also a lack of system-level design and validation in ThingML. 
CHESSIoT envisions having a consolidated and fully automated environment where users can combine both ThingML and CHESS technologies for industrial IoT systems development.   

\section{Proposed Approach}\label{sec:CHESSIoT}

\begin{figure}[t]
    \centering
    \includegraphics[width=\linewidth]{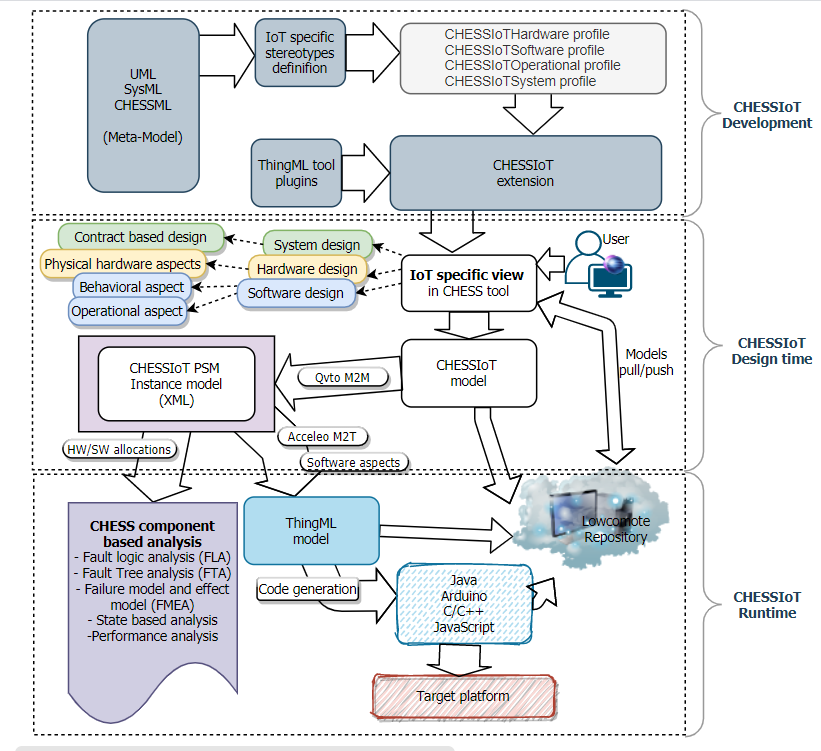}
    \caption{CHESSIoT approach}
    \label{fig:chessiotapproach}
    \vspace{-.6cm}
\end{figure}

The CHESSIoT tool has been developed on top of the recently released CHESS 1.0.0 tool \cite{CHESScomplex}. At design time, CHESSIoT enforces that all the components and elements be IoT specific and follows the already existing CHESS modelling methodologies. In particular, systems are specified in terms of a component-based design and employing a multi-view paradigm. The CHEESIoT extension consists of four different profiles depending on the specific view and need for the particular task at hand. As an addition to the domain-specific CHESS views, CHESSIoT adds the \textit{\textbf{IoT sub-view}} to permit the user to activate services and palettes related to the IoT domain. The CHESSIoT design extensions are available throughout the whole views provided by CHESS as long as the user activates the \textit{IoT sub-view} from the toolbar. Figure \ref{fig:chessiotapproach} shows the high-level representation of the CHESSIoT extension concerning its development infrastructure, design time, and run-time modelling functionalities. As shown at the top-right of Fig. \ref{fig:chessiotapproach} CHESSIoT consists of four profiles that are singularly described as follows.

\textbf{\textit{CHESSIoTSystem profile}}  serves to cover the system-level design aspects in which the systems main blocks and their interconnections are defined. This is done in the CHESS \textit{System View} invoking the \textit{IoT sub-view}. As an extension of the SysML blocks, the IoT high-level blocks and their corresponding flow-ports are defined and later annotated with formal properties as contracts. In this regard, the user benefits from the system level validation and verification, contract refinements, parameterized architecture, and trade-off analysis infrastructure, all provided by the CHESS tool. For the sake of the paper scope, we do not cover such aspects in this paper. 


The \textbf{\textit{CHESSIoTSoftware profile}} provides users with modelling constructs to describe the IoT software components and their behaviours. The software design is done in the \textit{Component view}. In this regard, the user can decompose the system's software components and sub-components provided by specific palettes. In CHESSIoT the component behaviours are defined by using state machines.  
    
\begin{figure}[h]
    \centering
    \includegraphics[width=\linewidth]{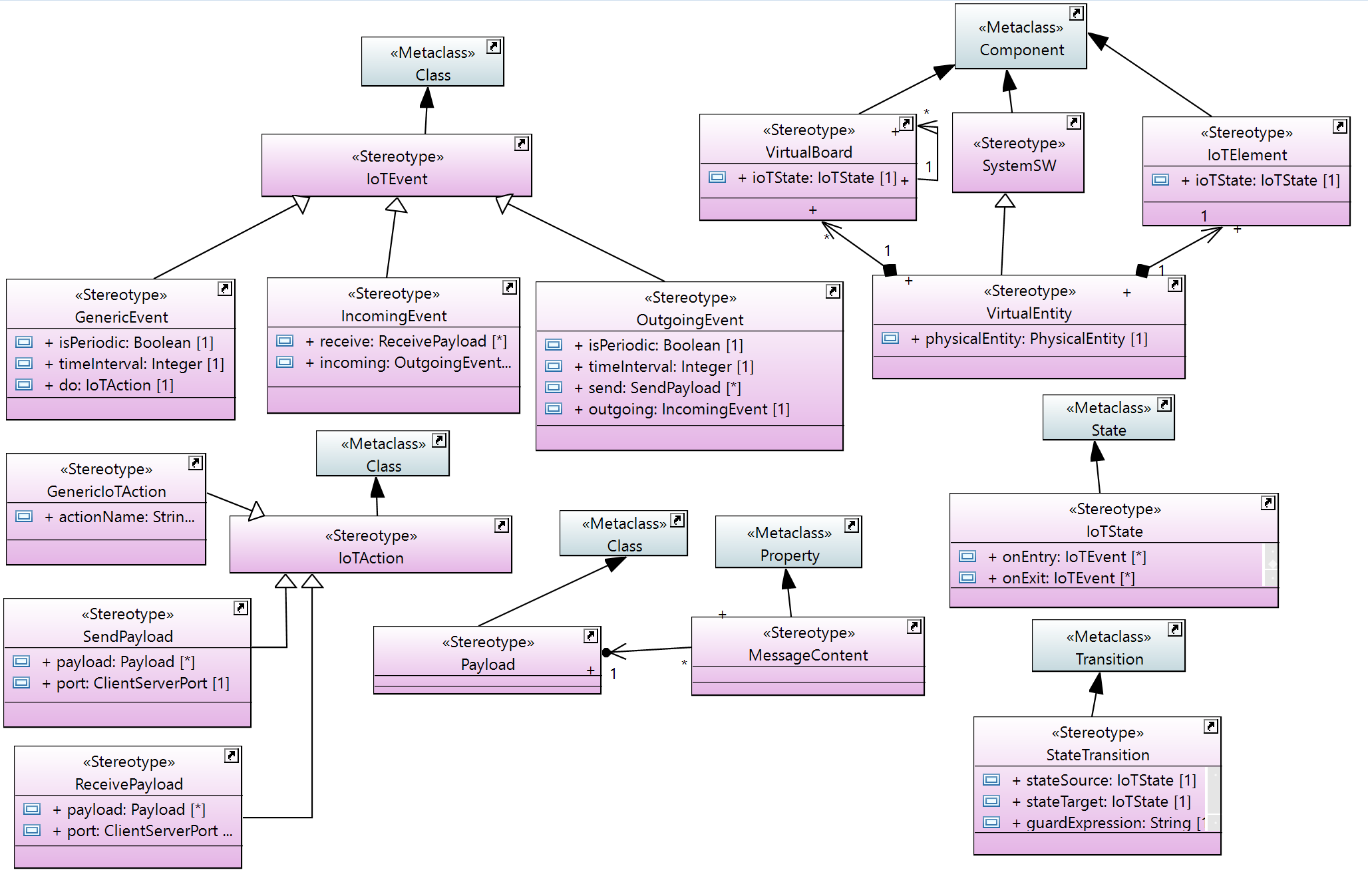}
    \caption{CHESSIoT software meta-model}
    \label{fig:chessiotsoftware}
    \vspace{-.3cm}
\end{figure}

Figure \ref{fig:chessiotsoftware} shows the CHESSIoT software profile meta-model, which consists of the following elements:
\begin{itemize}

    \item[--] The \textit{Virtual entity} permits the definition of digital representations of physical objects in which the systems resources are allocated. This entity consistently links to its corresponding "Physical Entity" to be defined in the deployment view when performing the hardware designs;
    
    \item[--] The \textit{Virtual board} enables the specification of IoT computing boards on which the application will run on. This component is a sub-component to \textit{VirtualEntity} and can be connected or have reference to one or many other virtual boards. The only crucial property it poses is its state which later gets defined;
    
    \item[--] The \textit{"IoTElement"} represents any other IoT Thing that comprises a system aside of being a \textit{VirtualEntity} or a \textit{VirtualBoard}. The internal software architecture can be specified using other internal sub-elements and the connectors and ports, while their behavioural specification is defined using custom states, events, and actions. Furthermore, the \textit{IoTElements} decomposed as IoT system sub-components are then connected to define the system's main components. To this level, the user can instantiate as many components as possible to meet any system complexity wanted. 
    
\end{itemize}    

In CHESSIoT, we have introduced a flexible event-based modelling mechanism. Normally, the UML state machine events and actions modelling process is complex and tricky when done using the normal Papyrus infrastructure. In our case, the \textit{IoTEvents} and \textit{IoTActions} can be modeled separately and later be invoked as many times as possible or by many different \textit{IoTElements} states.  An \textit{IoTState} always should be linked to an \textit{"OnEntry"} and an \textit{"OnExit"} events, which in turn can be either generic, incoming, or an outgoing event. The \textit{IoTAction} element is for sending and receiving payloads, and action types can then be associated with each event in the form of effect. A \textit{payload} is any type of message exchanged between \textit{IoTElements} through ports and can be reused as many times as possible. More information on this is also presented in Section \ref{sec:chessiot2thingml}.

\textbf{\textit{CHESSIoTHardware profile}} contains a physical deployment representation of the virtual components designed in \textit{Component view}. The hardware modelling activities are performed in \textit{Deployment View}. The hardware design also includes the definition of target platform specifications, such as the number of processors and core units. Most of the elements in this profile rely on MARTE \cite{MARTE}.

\textbf{\textit{CHESSIoTOperational profile}} contains all the information regarding the communication aspects of the system, and it extends the CHESS’s \textit{Component View}. Ideally, the information related to communication mode, servers, communication protocols, and storage resources will be modelled and analysed using this profile. The purpose of adding this part is to support the performance analysis of the involved resource blocks.

In the end, CHESSIoT also will allow the user to generate platform-specific code employing the ThingML platform automatically. Lastly, the user will interact with a remote repository by consuming an open API pushing or pulling artefacts. The high-level description of envisioned CHESSIoT to ThingML model transformation is described in the next section.

\subsection{The  CHESSIoT to ThingML model transformation} \label{sec:chessiot2thingml}

Modelling software component in CHESSIoT goes hand in hand with defining its behaviours, resulting in platform-specific code. The CHESSIoT component's semantics differs from the ThingML's to some extent, and that is why the mapping of the elements is needed to solicit an efficient transformation. In the following, we discuss how the different CHESSIoT modelling constructs contribute to the generation of target ThingML elements.

\textbf{A component}: In CHESSIoT, the software components such as \textit{IoTElement, VirtualBoard, VirtualEntity} are the main modeling elements. They are used to encapsulate the system's main part structure, operations, and behaviours. These components are mapped to ThingML's thing. 

\textbf{Provided/Required port}: Ports are used to support the communication between two or more components exposing or requiring the interfaces from other components. In CHESSIoT, the component's messages are passed through the port using the required or provided interface operations. During the transformation, the Required/Provided ports of the components are mapped to the \textbf{required/provided port} of a ThingML's thing.

\textbf{Operation}: In CHESSIoT, operations specify the functional behaviour of components. During the transformation, each component's operation is mapped to corresponding Thing's \textbf{function}. 

\textbf{Property}: Properties represent variable attributes local to a component. The property can be primitive or be an instance of other components. Same as in ThingML, properties are used to retain the variable functional value of a Thing, in which during the transformation, the component's property will be mapped to thing's \textbf{property}.

\textbf{Payload}: This is a standalone and straightforward object to carry information to be passed between components. The payload will be mapped to a \textbf{Message} in the ThingML model. In CHESSIoT, the payload can have zero or many primitive or derived properties to be defined in a message. For instance, suppose a component message to be communicated among components contains a string value, an integer or even an instance of another payload. In this case, a payload will include three different attributes, which will be represented as message arguments in ThingML.

\textbf{IoTState}: This serves to keep the component state from its initial participation until its disposal in the system. \textit{IoTState} extends actual UML states but in addition to that, \textit{IoTState} carries information related to what events need to be taken care of at a certain point in time. For instance, \textit{OnEntry} or \textit{OnExit} events are triggered when entering or exiting a state. An IoT state can also trigger an internal event, which corresponds to an internal action to be taken. During the transformation, the IoTState will be mapped to the \textit{ThingML state}, same goes to \textbf{State transition} that will also be mapped to their corresponding transition provided by ThingML. 

\textbf{IoTEvent}: Events in CHESSIoT are triggered in a different manner depending on the state of the component. An \textit{IoTvent} can be incoming, outgoing, or generic, which means it can come from the inside-out, from the outside, or internal. A \textit{GenericEvent} is an event that gets triggered internally to the component, for example, changing variable value.  CHESSIoT events are mapped to corresponding ThingML \textit{event(s)}.

\textbf{IoTAction}: IoTAction(s) can be of different types depending on the kind of action to be performed. For instance, the \textit{SendPayload} action is referred to when an \textit{OutgoingEvent} is triggered to send the payload through a specified port while \textit{ReceivePayload} is used on \textit{IncomingEvent}s to receive messages from another components. A \textit{GenericAction} does not require to access the component's ports, for example, changing the component's property value. During the transformation, \textit{IoTAction}s will be mapped to  corresponding ThingML actions.

\textbf{State Transition}: In CHESSIoT, state transitions enable transiting from the source state to a target state, abiding the trigger from the guard value. Guard expressions are boolean expressions defined based on state values. They serve to initiate a state transition by checking whether the \textit{OnExit} event has been performed correctly. During the transformation, \textit{State transition}s will be mapped to the corresponding transitions in ThingML.

\begin{table}[h]
  \caption{Proposed CHESSIoT2ThingML mapping}
  \label{tab:extension}
  \begin{tabular}{ccl}
    \toprule
    \textbf{CHESSIoT element } && \textbf{ThingML element}\\
    \midrule
    \texttt{Component} && Thing\\
    \texttt{Provided/required port} &&  Provided/required port\\
    \texttt{Operation} && Function\\
    \texttt{Property} && Property\\
    \texttt{Payload} && Message\\
    \texttt{IoTState/Transition} && State/Transition\\
    \texttt{StateGuards} && Guards\\
    \texttt{IoTEvent/Action} && Event/Action\\
    \bottomrule
  \end{tabular}
\end{table}

\section{A real-time safety use case in the Industrial IoT domain}  \label{sec:usecase}


In this section, an explanatory industrial safety use case is modelled and analysed employing the proposed approach. The CHESS tool offers several analyses like dependability and real-time schedulability analyses, which are enabled once the model's component instances are annotated with extra-functional properties.  In this section, we cover only real-time schedulability analysis. 

\begin{figure}[t!]
    \centering
    \includegraphics[width=\linewidth]{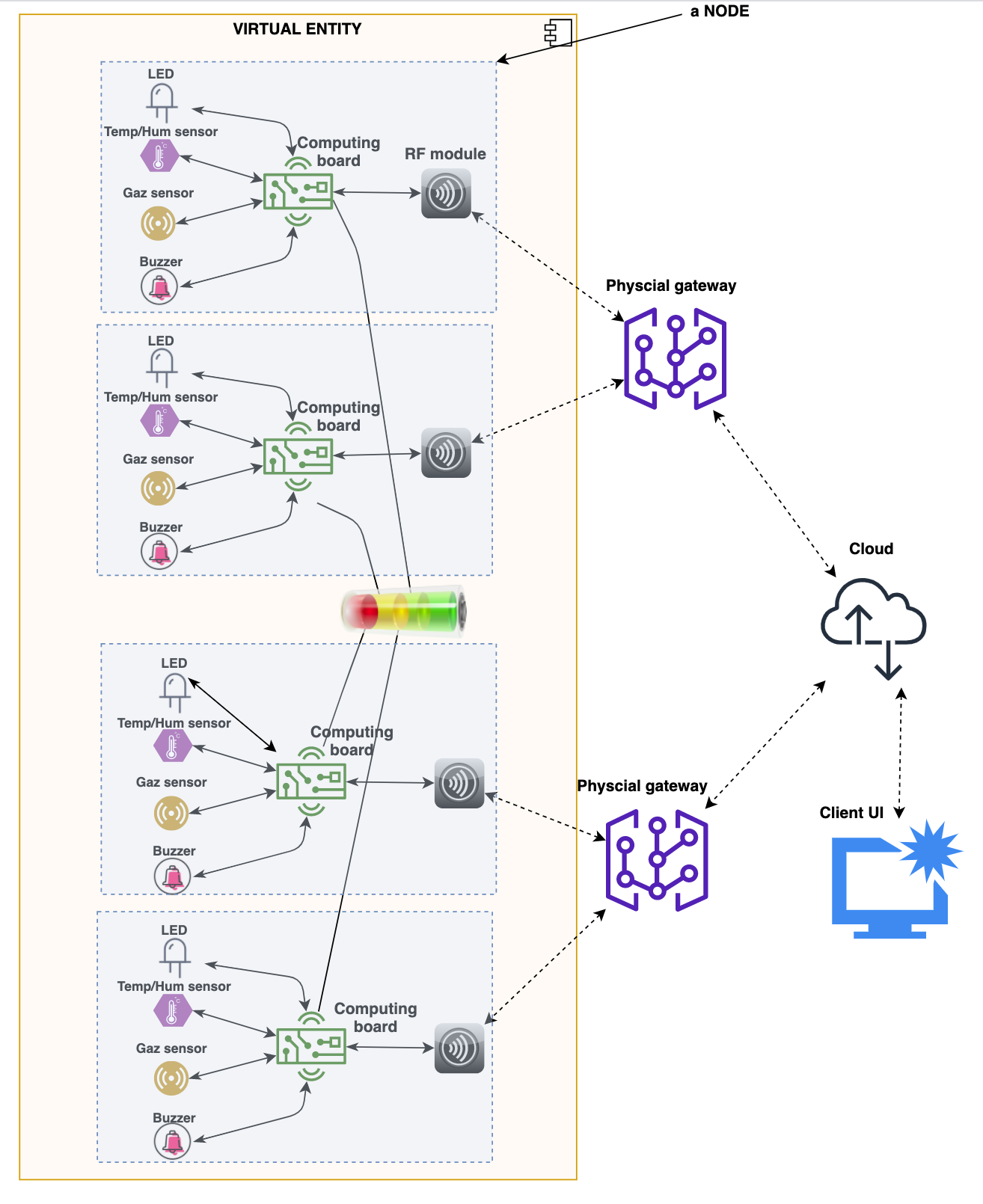}
    \caption{Physical structure of the simple IoT system}
    \label{fig:safetyUseCase}
    \vspace{-.3cm}
\end{figure}

Safety in Industrial IoT systems is crucial to ensure the reliability of the system and the security of the people. Current systems impose the usage of as early detection of any potential safety-related intrusion. When not taken care of, it might bring an immense loss of assets, and in some cases, might even cost human lives as well. A few modern IoT framework configuration approaches, for example \cite{uml4iot}, have been proposed for monitoring events and act upon a given exceptional case. In the example shown in Figure \ref{fig:safetyUseCase}, a basic safety system is proposed. The system objective is to gather information through specific nodes deployed into a modern environment to collect environment data, and in case of event an unprecedented issue, the system triggers an alarming mechanism to limit further damages.

The system consists of different nodes. The deployed node runs on a computing device such as \textit{Arduino  Uno}\footnote{\url{https://store.arduino.cc/arduino-uno-rev3}}. Humidity and temperature data are collected using attached \textit{DHT-11 temperature and humidity sensor}\footnote{\url{https://components101.com/sensors/dht11-temperature-sensor}}, whereas the gas level data are collected  through the attached \textit{MQ2-type gas sensor}\footnote{\url{https://components101.com/articles/introduction-to-gas-sensors-types-working-and-applications}} to detect flammable and combustible gasses. Collected data is directly sent to the remote gateway wirelessly using the attached \textit{HC-12 Wireless Transceiver}\footnote{\url{https://www.elecrow.com/download/HC-12.pdf}}. The system triggers a red light-emitting diode (LED) and sounds an electromechanical alarm in case of any unusual detection. Normally, this system can be used in different industrial contexts or even at home for safety purposes.  

In this example, we do not cover the communication layer and the cloud-side designs. This is deemed to be done using the operational profile, which is currently yet fully developed. Besides that, the analyses being performed will be one on the ThingLayer components. More on this is discussed in Section \ref{sec:realtimeanalysis}. In the following, we present the modelling phases of the proposed system.

\subsection{Modeling software components}

The first step of the proposed approach consists of modelling the node's main components, corresponding types, and the interfaces they implement. We suppose that the system requirements have been modeled in the CHESS's requirement view. Modelling the software constructs is done in the component view. Following the approach presented before, the \textit{computing board} in CHESSIoT is defined as a \textit{Virtual board}, other devices such as LEDs, sensors, and buzzer are defined as \textit{IoTElements}. As CHESSIoT follows a component-based approach, only one component needs to be defined, and it can be instantiated and reused as many times as possible.

A \textit{virtual entity} is considered as a physical object where the computing node will be deployed. In other words, they won't implement any interface or perform any action. As described before, the interfaces contain functional operations to be carried out during the communication, and they automatically get applied to the elements that implement them.

\begin{figure}[t]
    \centering
    \includegraphics[width=\linewidth]{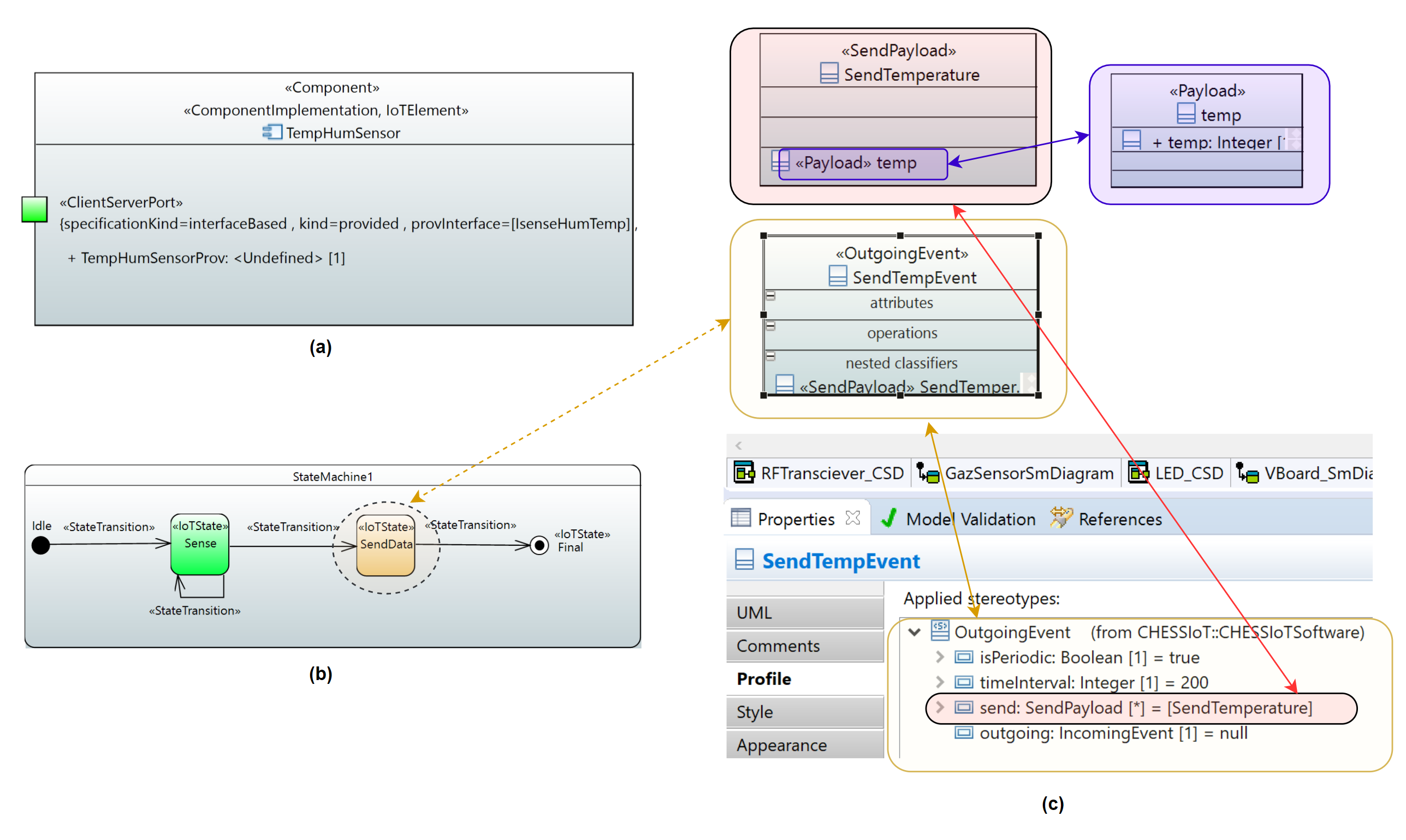}
    \caption{\textit{(a)Temp/hum sensor internal structure, \\(b) State machine, (c) Event, action and payload definition}}
    \label{fig:humtemp}
    \vspace{-.6cm}
\end{figure}

After defining \textit{Component}, \textit{ComponentType}, \textit{Interfaces} elements and their corresponding \textit{Operations}, each component's internal structure is decomposed. This is done by using the UML composite structure diagram. \textit{IoTElemnts} such as LED, sensors, and buzzer are considered as sub-components of the \textit{Virtual board}. The internal structure is specified by defining component's required/provided ports with the corresponding interfaces they expose or require. At this stage, CHESSIoT allows modelling the component behavioural aspects using the UML state machine. The component's event, action, and payloads are modeled using the component's inner class diagram and then linked back to their corresponding states. Note that the \textit{IoT sub-view} element provided by CHESSIoT needs to be opened here to have the right palette containing the behavioural resources. Figure \ref{fig:humtemp} shows the three different modelling parts of the internal structure of a temperature/humidity sensor. 

\begin{figure}[t]
    \centering
    \includegraphics[width=\linewidth]{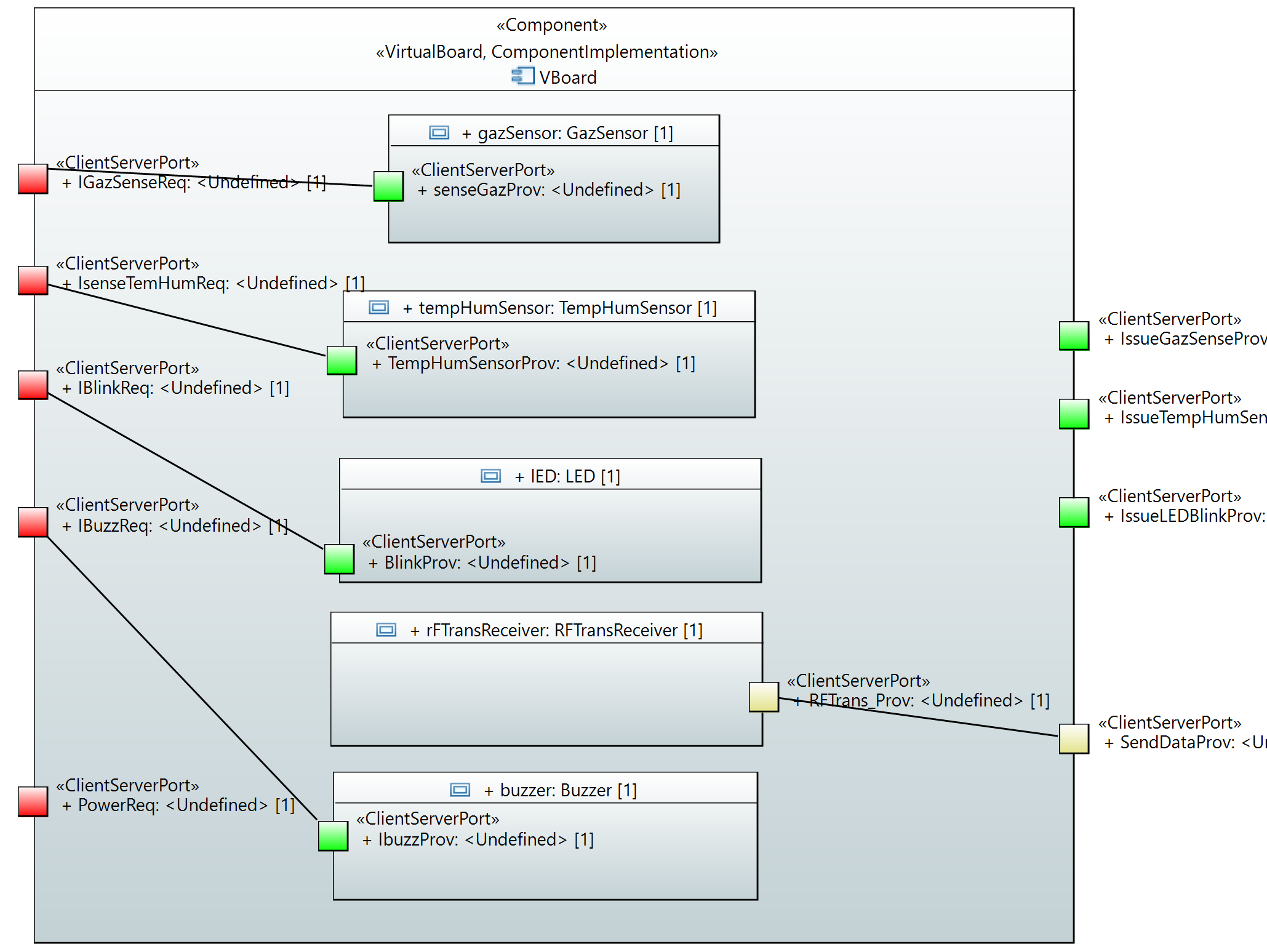}
    \vspace{-.2cm}
    \caption{Virtual board internal structure}
    \vspace{-.6cm}
    \label{fig:virtualBoard}
\end{figure}

As shown in Fig. \ref{fig:humtemp}, temperature/humidity sensors have only one port which provide the \textit{IsenseHumTemp} interface. The behavioural modelling of it is mainly for code generation purposes. This process has to be done for each defined \textit{IoTElements}. The next step is to model the node, which is a \textit{Virtual board}.  At this level, the \textit{IoTElement} can be instantiated as many times as possible to achieve the desired structure of the node. The internal structure of the virtual board can be modeled as shown in Fig. \ref{fig:virtualBoard}. The virtual board also contains required (in red) and provided (in green) ports. The provided ports of the IoTElements have to be connected to their corresponding required ports of the board and vice-versa. The ports in yellow represent a port that provides and require an interface. This can apply to the data communication between the \textit{RFModule} and the computing board.

The same as the other elements, the board's behavior specification has to be defined too, but we won't present it for the sake of space. 
The virtual entity comprises as many virtual board instances as possible and one or more power source instances. 
In our case, we used four board instances with one power source.    

The next stage involves specifying different physical characteristics of the target platform, such as the number of processors and number of cores. This part extends the MARTE profile typically and this is done in CHESS's deployment view. In our case, the system will be deployed to one \textit{Physical entity} which will comprise four different processors. Three of the processors will run on one core each, while the fourth runs on two cores. This process is performed in \textit{Deployment view}. At this stage, we generated the software and hardware instances that contain the containers and connectors representing components instances and their connectors.    


\subsection{Real-time schedulability analysis} \label{sec:realtimeanalysis}

CHESS tool offers the means to perform real-time analyses on the modelled instance model. Schedulability analysis is performed employing MAST \cite{MAST}, a timing analysis tool that relies on the system's component timing requirements. To perform such analysis, the user needs to annotate the real-time temporal logic properties on each component's operations. Those properties include the timing request type (i.e., periodic or sporadic), the worst-case execution time of a request, the priority, and the desired execution deadline. This activity is performed in the CHESS's \textit{InstanceView}. 

CHESS also supports the software to hardware component allocation, which includes allocating component to processor cores. In our case, each node is allocated to one processor running on a single core, except for the second processor, which had two cores and can accommodate two nodes. For analysis purposes, we have allocated the battery unit to one processor running on one core also.

\begin{figure}[t!]
\minipage{0.48\linewidth}
  \includegraphics[width=\textwidth]{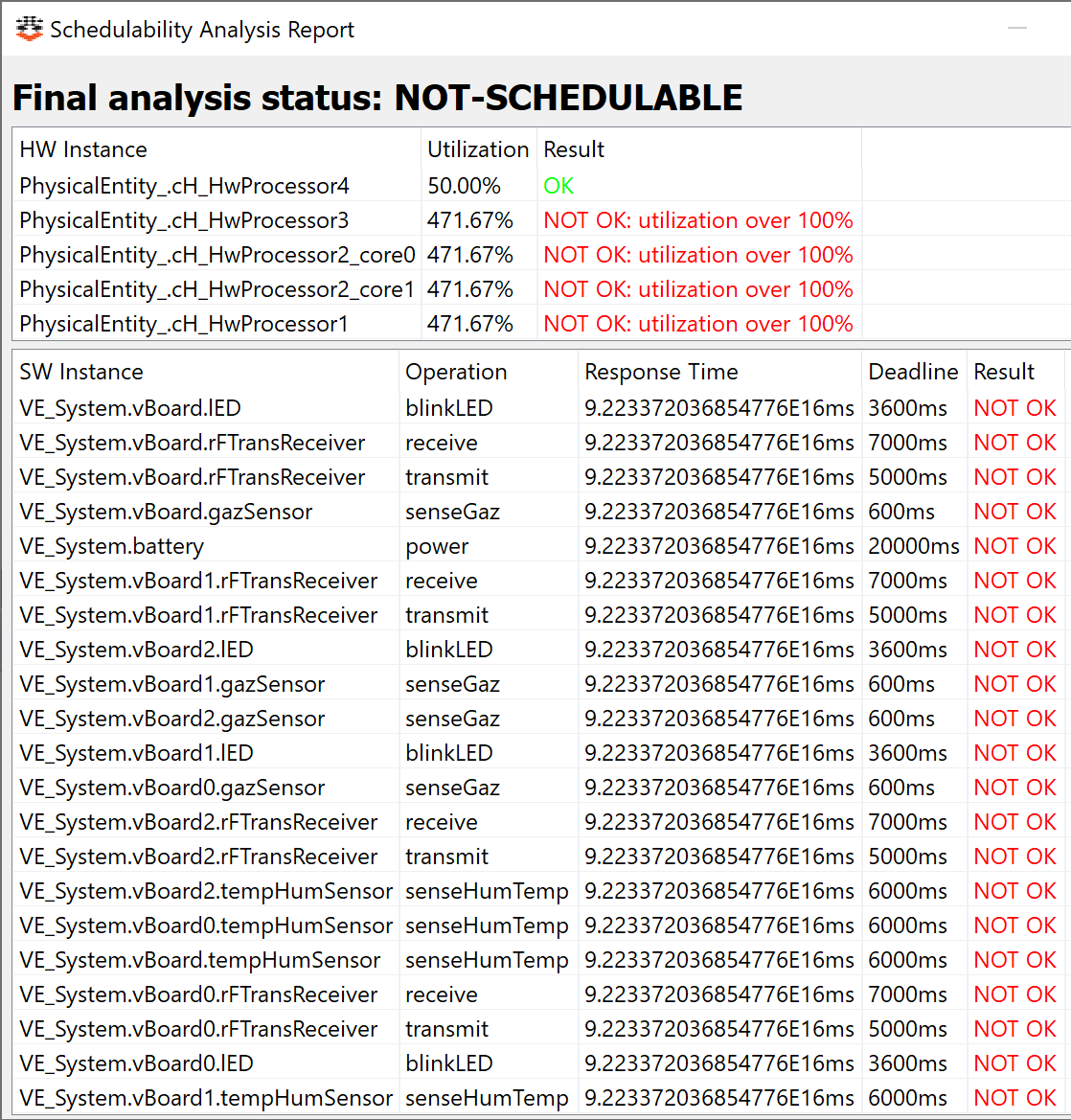}
  \caption{Schedulability\\
  results (NOT OK)}\label{fig:not_scheddulable}
\endminipage\hfill
\minipage{0.48\linewidth}
  \includegraphics[width=\textwidth]{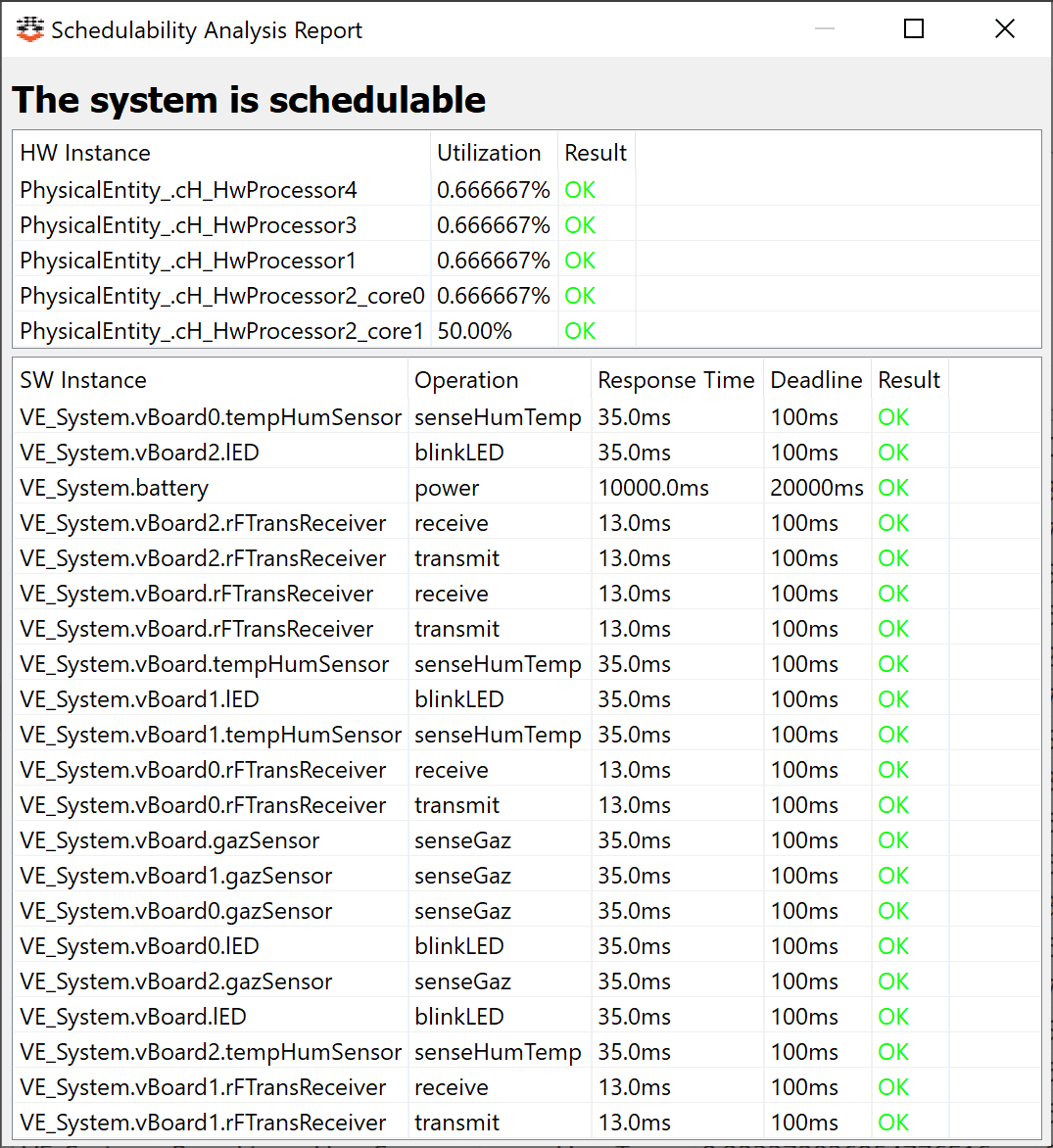}
  \caption{Schedulability \\ results ( OK)}\label{fig:schedulable}
\endminipage\hfill
\vspace{-.3cm}
\end{figure}

In the \textit{Analysis View}, we can now specify the analysis context, which will contain the software-hardware instance specifications to run the schedulability analysis. The schedulability results indicated in Fig. \ref{fig:not_scheddulable} show that the system cannot be schedulable due to the excess memory utilization according to the specified real-time properties. We can also look at the timing deadline constraint specified is a way less than the response time. In this case, when the deadline is increased, the execution time also increases, amplifying the response time exponentially as more components are still waiting to respond to a given request. 
To make the system schedulable, we have reduced the worst-case execution time by 70\% and fixed the deadline for all operations to 100milliseconds. We have also separated the nodes from running on the same processor and deploy the battery unit to the second processor's second core. As presented in Fig. \ref{fig:schedulable}, the system has now become schedulable. 

Given the schedulability information presented above, we can now know how we may go ahead with the real-case deployment e.g., by having a clear picture of processor types or core counts. Another helpful information we can account too is the event deadlines specifications. In our example, the \textit{SendPayload} action is periodic, and it is deemed to be sent every 200ms (refer to figure \ref{fig:humtemp}), which is perfectly fine because each action in the system needs an average of 100ms to be schedulable.

\section{Related Work}\label{sec:related}

Modelling and analysis of the Internet of Things is not a new topic in general, several approaches have already been studied and validated, but few of them focus on the Industrial domain. This section will look briefly at some related research to our system and discuss their differences and correspondences. For the sake of the topic of interest, we will only cover the MDE approaches that extend uses Eclipse Papyrus \footnote{\url{https://www.eclipse.org/papyrus/}} modelling as an underlying environment.

In \cite{chessIoT}, the CHESS tool has been used to model the life-cycle of scalable and distributed intelligent IoT applications. Although this was a great start towards using CHESS for modeling IoT systems, it is generic considering the specificity of IoT applications. In our approach, we emphasise decoupling the modelling environment for IoT elements to the whole CHESS platform. 

In \cite{COMFIT} the authors presented a Cloud and Model-based IDE for the Internet of Things tool (COMFIT) to target the wireless sensor networks (WSN) applications for IoT. The COMFIT modelling environment is built on top of Papyrus, and it presents a simple multi-view environment to model the system's requirement, structural and behavioural aspects. The wireless nodes of the system and their communication links are created in the structural view, while the model activities and behaviours are modelled as functional units, which later get linked according to the desired execution sequence. Finally, the tool provides the model checking infrastructure respecting the OCL rules specified in the meta-model. 



Ciccozzi et al. presented the MDE4IoT tool \cite{MDE4IoT} developed to support the modelling of self-adaptation of connected Emergent Configurations  (ECs) referred to as Things in the IoT domain.  The authors present a simple smart street light case to highlight and validate their approach characterized by the multi-view and separation of concern capability of the MDE4IoT tool.
The run-time adaptations are meant to be performed automatically by specific in-place model transformations that modify the source models. The generation of code from UML state machines has been addressed, whereas the fUML’s ALF complex execution behaviours of the system’s components have been only described. Although this approach is closely related to CHESSIoT, the paper still does not address any analysis aspects of the system.

In \cite {papyrus4iot} the authors introduced Papyrus4IoT, a modelling tool developed under the S3P project. The authors presented a modelling methodology that uses the UML use case diagrams and early system conceptual system specification requirements. Later the designer can define process specification definition, functional, operational platform, and finally, the deployment is done by allocating the system's functional blocks to the device processing units. Unlike \cite {papyrus4iot}, our emphasis is given on separation of concerns and supporting analysis that has not been addressed in \cite {papyrus4iot}. 


Conzon et al. \cite{brainIoT} have presented the BRAIN-IoT framework, an integrated modelling tool to ease rapid prototyping of intelligent cooperative IoT systems based on shared models. The constructed models are transformed into XML format before being uploaded to the BRAIN-IoT marketplace for future reuse. The model and its manifest containing the metadata about the software are passed into the BRAIN-IoT code generator which automatically generates the OSGi artifacts which are also get stored in the BRAIN-IoT marketplace.

In \cite{uml4iot}, the authors introduced UML4IoT domain-specific modelling language to tackle the Industrial Automation Thing (IAT) domain. In UML4IoT, the system's components are transformed into IATs by IoTwrapper and later integrated into the IoT-based industrial automation environment. The RESTful paradigm is adopted for enhancing the connectivity with third-parties resources. Although their approach focuses more on the industrial use case, it differs from CHESSIoT concerning the multi-view and analysis support. 

Authors in \cite{sysml4IoT} introduced a SySML4IoT modelling and analysis platform derived from SysML to support the IoT domain. Following the IoT-A reference architectural reference in \cite{enablingIoT}, the authors introduced the SysML2NuSMV translator. 
The tool has later been extended by \cite{sysml4IoTExtend} to support the design, and the usage of the public/subscribe paradigm to model the communication relationships with other systems. Although the authors suggest the system analysis by considering only the quality of service through model checking, we see it as not sufficient in the industrial IoT domain.

From the above discussion, we can see that most of the proposed UML based modelling approaches for industrial system focus more on design and code generation; we see a significant lack in the analysis mechanism of IoT systems in general with the industrial case in particular. We think that this is an excellent direction for CHESSIoT to explore to address such challenges.

\section{Conclusion and future work}\label{sec:conclusion}

Developing Industrial IoT systems has to cope with several challenges, ranging from the heterogeneity in different players to the types of transferred messages. This paper has proposed the CHESSIoT extension covering the Industrial IoT domain on top of the already existing CHESS tool. CHESS is known for its success in modelling and analysis of industrial system engineering applications.  We have presented an envisioned mapping from the CHESSIoT model to the ThingML model to support the code generation later. Finally, we showed a modelling and analysis example of a real-time safety use case to validate the current capabilities of the extension.  

To enhance the scalability of our approach, in our plans, we would like to explore to possibilities of combining the graphical modelling with a textual-based interface. We would also want to extend different real-time and dependability analyses provided by CHESS to cover the CHESSIoT approach taking care of IoT-specific aspects in the analysis processes. ThingML platform is compelling, but as we have seen, there are still other essential aspects it can't cover now, for example, concerning model checking and analysis. In our plans, we would like to fully automate the ThingML code generation process from CHESS, in which the platform-specific code generation will be complied and generated directly from the CHESSIoT environment. In our future, we would also like to explore the possibility of exposing the analysis infrastructure so that any external user can consume the tool remotely by consuming an open API.

\begin{acknowledgments}
This work has received funding from the Lowcomote project under European Union’s Horizon 2020 research and innovation program under the Marie Skłodowska-Curie grant agreement n\si{\degree} 813884. 
\end{acknowledgments}

\bibliography{Bibliography}
\end{document}